\documentclass{emulateapj}
\pdfoutput=1 
\usepackage{amsmath,amstext}
\usepackage[T1]{fontenc}
\usepackage{graphicx}
\usepackage{epstopdf}
\usepackage{amsmath}
\usepackage{hyperref}
\usepackage{cleveref}
\usepackage{xcolor}

\shorttitle{An Unusual H-poor Superluminous Supernova: SN2017egm}
\shortauthors{Yan, Lin et al.}

\begin{document}

\title{Far-UV HST Spectroscopy of An Unusual Hydrogen Poor Superluminous Supernova: SN2017egm}

\author{Lin Yan$^{1}$, D. A. Perley$^2$, A. De Cia$^3$, R. Quimby$^4$, R. Lunnan$^5$, Kate H. R. Rubin$^4$, P. J. Brown$^6$}
\affiliation{$^1$Caltech Optical Observatories and IPAC, California Institute of Technology, Pasadena, CA 91125, USA; email: lyan@caltech.edu}
\affiliation{$^2$Astrophysics Research Institute, Liverpool John Moores University, IC2, Liverpool Science Park, 146 Brownlow Hill, Liverpool L3 5RF, UK}
\affiliation{$^3$European Southern Observatory, Karl-Schwarzschild Str. 2, 85748 Garching bei Muenchen, Germany}
\affiliation{$^4$ Department of Astronomy, San Diego State University, San Diego, CA 92182, USA}
\affiliation{$^5$ Oskar Klein Center, Astronomy Department, Stockholm University, SE-106 91 Stockholm, Sweden}
\affiliation{$^6$George P. and Cynthia Woods Mitchell Institute for Fundamental Physics \&\ Astronomy, Texas A. \&\ M. University, Department of Physics and Astronomy, 4242 TAMU, College Station, TX 77843, USA}


\begin{abstract}

SN2017egm is the closest ($z=0.03$) H-poor superluminous supernova (SLSN-I) detected to date, and a rare example of an SLSN-I in a massive, metal-rich galaxy. We present the {\it HST} UV \&\ optical spectra covering $1000 - 5500$\,\AA\ taken at +3\,day relative to the peak. Our data reveal two absorption systems at redshifts matching the host galaxy NGC3191 ($z=0.0307$) and its companion galaxy ($z=0.0299$) 73$\arcsec$ apart. Weakly damped Lyman-$\alpha$ absorption lines are detected at these two redshifts, with HI column densities of $(3.0\pm0.8)\times10^{19}$ and $(3.7\pm0.9)\times10^{19}$\,cm$^{-2}$ respectively. This is an order of magnitude smaller than HI column densities in the disks of nearby galaxies ($>10^{10}M_\odot$) and suggests that SN2017egm is on the near side of NGC\,3191 and has a low host extinction ($\rm E(B-V)$\,$\sim$\,0.007).  Using unsaturated metal absorption lines, we find that the host of SN2017egm probably has a solar or higher metallicity and is unlikely to be a dwarf companion to NGC\,3191. Comparison of early-time UV spectra of SN2017egm, Gaia16apd, iPTF13ajg and PTF12dam finds that the continuum at $\lambda$\,$>$\,2800\,\AA\ is well fit by a blackbody, whereas the continuum at $\lambda$\,$<$\,2800\,\AA\ is considerably below the model. The degree of UV suppression varies from source to source, with the 1400\AA\ to 2800\AA\ continuum flux ratio of 1.5 for Gaia16apd and 0.4 for iPTF13ajg.  This can not be explained by the differences in magnetar power or blackbody temperature. Finally, the UV spectra reveal a common set of seven broad  absorption features and their equivalent widths are similar (within a factor of 2) among the four events.

\end{abstract}


\section{Introduction}

The spectra of hydrogen-poor superluminous supernovae (SLSNe-I) peak, and show their most distinctive spectroscopic features, in the ultraviolet \citep{Quimby2011}.  These wavelengths are inaccessible from the ground, except at high redshift where the vast distances limit the S/N that can be achieved.  Space-based UV spectra can cover these features for nearby SNe, but the limited sensitivity of UV spectrographs (even \emph{HST}) restricts observations to very nearby and rare events.  Up until now only one high-quality far-UV spectrum of a SLSN has been taken (Gaia16apd at $z=0.102$; \citealt{Yan2017}).

Knowledge of the FUV spectra of SLSNe is important for several reasons: not only to investigate the structure and composition of the photosphere, but also to improve modeling of SLSN light curves and measure the bolometric energy release.  Physical parameters are usually derived by fitting bolometric light curves (LC), estimated from broadband optical photometry in two or three filters. One of the significant uncertainties in this process is bolometric correction, especially at early times when as much as 50\%\ of the bolometric luminosity is likely to escape in the UV.  {\color{black} Many} previously published studies have relied on either crude pseudo-bolometric LCs (summing up the available optical bands), or by assuming a blackbody spectral energy distribution (SED), even though these events are known to have broad, deep UV absorption features \citep{Quimby2011}.   

{\color{black}  More recently, \citealt{Nicholl2017b, DeCia2017b} compute magnetar model parameters for a large sample of SLSNe-I by applying a single SED template to broad band light curves.  This SED template  is derived for the purpose of fitting broad band photometry. It does not match well with the {\it spectral continuum} of Gaia16apd. Broad band SEDs are affected by both continuum and broad, prominent absorption features \citep{Yan2017}.  The focus of this paper is new in its investigation of how UV spectra vary among different SLSN-I, including both spectral continuum and absorption features.} 

SN2017egm (Gaia17biu) exploded in NGC\,3191 at a distance of 140\,Mpc ($z=0.030$), making it the closest known SLSN-I by a large margin \citep{Delgado2017, Dong2017}.  This offers an excellent opportunity to study the UV spectrum of an SLSN-I in detail, and to examine the diversity of this class at these wavelengths. 

A particularly notable aspect of SN2017egm beyond its sheer proximity is its unusual location: the host is a large spiral galaxy with stellar mass $M_{*} \sim 10^{10.7}M_\odot$ and metallicity $Z \sim (1.3-2)Z_\odot$ \citep{Nicholl2017,Bose2017}.  This seems to argue against the consensus from several previous studies that almost all SLSNe-I occur in dwarf host galaxies with low metal abundances ($M_{star} \le 10^{9.7}M_\odot$ and $Z \le \sim 0.4Z_\odot$) \citep{Lunnan2014, Leloudas2015, Perley2016, Schulze2016, Chen2017}.  SLSNe-I have been observed in high-mass hosts before, but only far from the center, and it has been argued that the metallicity at the SLSN site could still be low in these cases \citep{Perley2016}.  The projected position of SN2017egm is 
very close to the host nucleus, and archival integral field unit (IFU) spectra of NGC\,3191 make it possible to zoom in onto the spiral arm region where the supernova exploded, and measure the nebular gas-phase metallicities with a resolution of $\sim$1 kpc. Two independent groups have found conflicting results, one with solar to super-solar metallicity \citep{Chen2017b}, and the other sub-solar metallicity \citep{Izzo2017}.  

In this paper we present our HST far- and near-UV spectra of SN2017egm, and compare this event with the only pre-existing FUV SLSN spectrum (Gaia16apd) as well as two other high-quality near-UV SLSN spectra.  We also exploit the narrow absorption transitions seen in our spectra to constrain the kinematics and chemical abundances of the system, providing confirmation (independent of emission-based methods) that the SN occurred along a metal-rich sightline and suggesting that SLSN production is not fully suppressed at high-metallicity.

\section{HST Data}
SN2017egm was first discovered as a transient on 2017 May 23 UT by the Gaia Photometric Survey \citep{Gaia2016, Delgado2017}. Optical spectra taken on May 30, 2017 UT by \citet{Dong2017} classified this event as a SLSN-I.  We immediately triggered our approved {\it HST} ToO program (PID: 14674, Quimby), using both the Cosmic Origin Spectrograph (COS) and the Space Telescope Imaging Spectrograph (STIS) on board {\it HST}.  Our single-epoch UV spectrum was obtained 2017 June 23 (MJD = 57927\,d); the estimated peak and explosion dates are MJD = 57924 and 57890\,  respectively \citep{Nicholl2017}, leading to the time lag between the {\it HST} UV spectra and the explosion date, $\Delta t_{exp}$ of 35.9\,day (rest-frame).  As listed in Table~\ref{obslog}, the COS and STIS spectra were taken with G140L, G230L and G430L gratings with moderate spectral resolutions. At 1500\AA\, the {\it HST} COS spectra have a velocity FWHM of $\sim 120$\,km\,s$^{-1}$. We use the reduced spectra provided by the {\it HST} pipelines. 

Our analyses also consider three other SLSNe-I, which have UV spectra before or near maximum light. The Gaia16apd {\it HST} spectra were taken with the similar setup as that of SN2017egm except G430L grating, and have been published in \citet{Yan2017}. PTF12dam was observed with the {\it HST/WFC3} slitless grism G280 at $-18.4$\,days before the peak, and $+44.2$\,days after the explosion ({\color{black} \citealt{Nicholl2013};\citealt{Vreeswijk2017}};  Quimby et al. 2017). As shown in Table~\ref{tab:date}, PTF12dam has the longest time lag between the UV spectrum observation and the explosion date. Its near-UV {\it HST} spectrum reaches out to 1850\AA\ and has a low resolution of 70 at 3000\AA\ (Quimby et al. 2017).  The iPTF13ajg spectrum was taken with X-shooter on the VLT telescope on 2013-04-17 at $-4$\,days from the maximum light. With $z\sim0.7$, this spectrum is the best quality near-UV spectrum taken with a ground-based telescope for a SLSN-I \citep{Vreeswijk2014}.  

To date, there are {\it HST} spectra for only five SLSNe-I, including SN2017egm, Gaia16apd, PTF12dam, PTF11rks and SN2011ke. The spectrum for SN2011ke is taken at late phase (+24\,day post peak), and the data for PTF11rks has poor SNR (Quimby et al. 2017).  This paper focuses on the pre-peak UV spectra, therefore, does not consider SN2011ke and PTF11rks data. Optical spectra of SLSNe-I at $z>0.5$ from ground-based telescopes also reach the rest-frame UV. However, most spectra have low SNR and their continua are not well characterized. The exception is iPTF13ajg which has a high quality optical spectrum near the maximum light.

\section{Results \label{sec_result}}

\subsection{The Diversity and uniformity of SLSN-I UV spectra at maximum light}
With these early-time UV spectra, we have two goals -- (1) quantify SLSN-I {\color{black} spectral continuum and absorption feature} variations, which is important for estimating bolometric corrections and for modeling light curves of SLSNe-I based on broad band optical photometry, (2) identify UV, especially far-UV, spectral features. 

\begin{figure}[t!]
\includegraphics[width=0.5\textwidth]{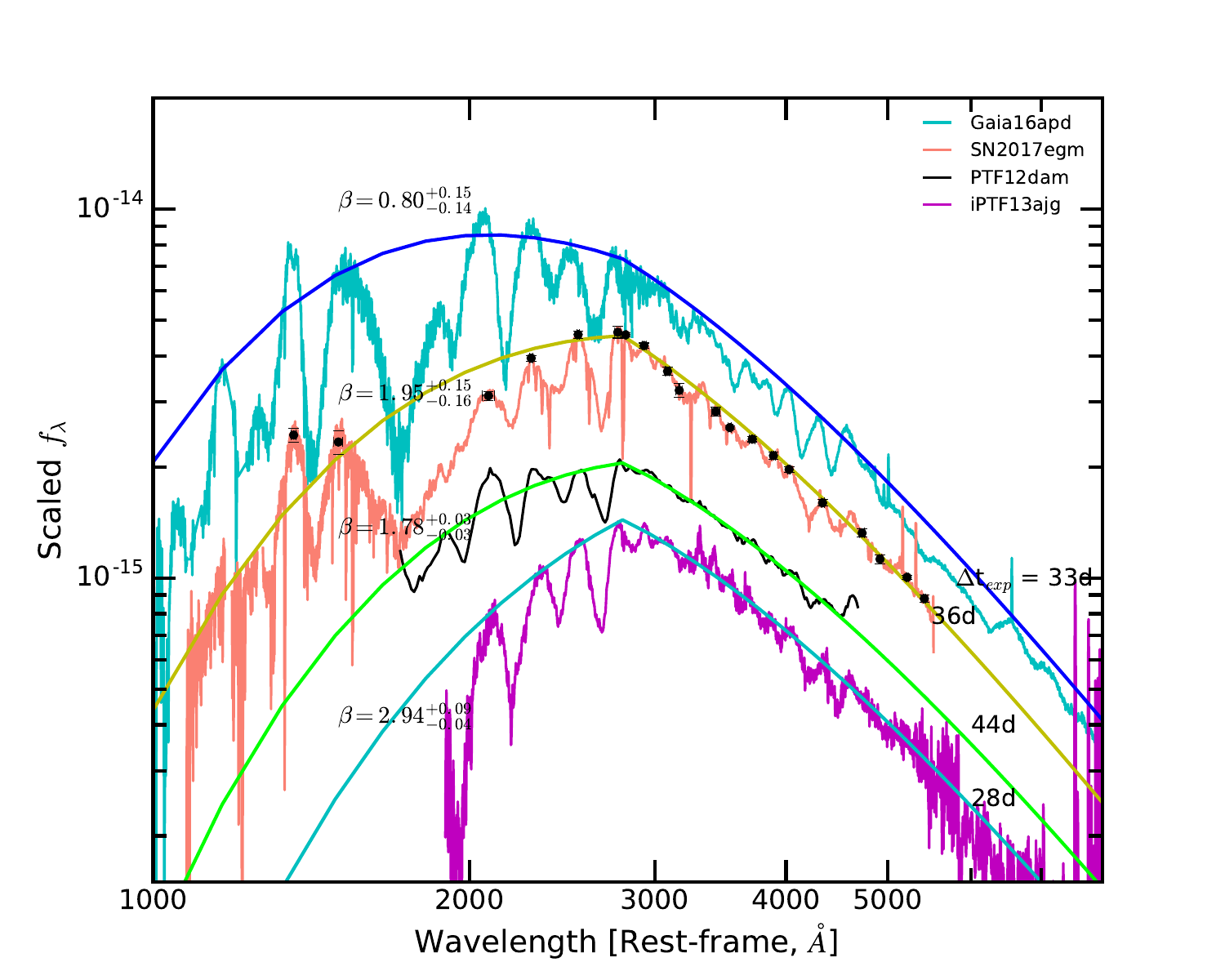}
\caption{ The UV spectra of four low-$z$ SLSNe-I. The Y-axis is flux density in erg/s/cm$^2$/\AA. The spectra are scaled to offset them from each other for visual clarity.  We also show one example of how we select spectral regions (solid black points for SN2017egm) as the input to the MCMC calculation. MCMC model fits to the continua are overlaid based on the parameters derived from the MCMC calculations with the fixed $\lambda_0$ and listed in Table~\ref{mcmc2}.  \label{fig:uvspec}}
\end{figure}

\begin{figure}[t!]
\includegraphics[width=0.5\textwidth]{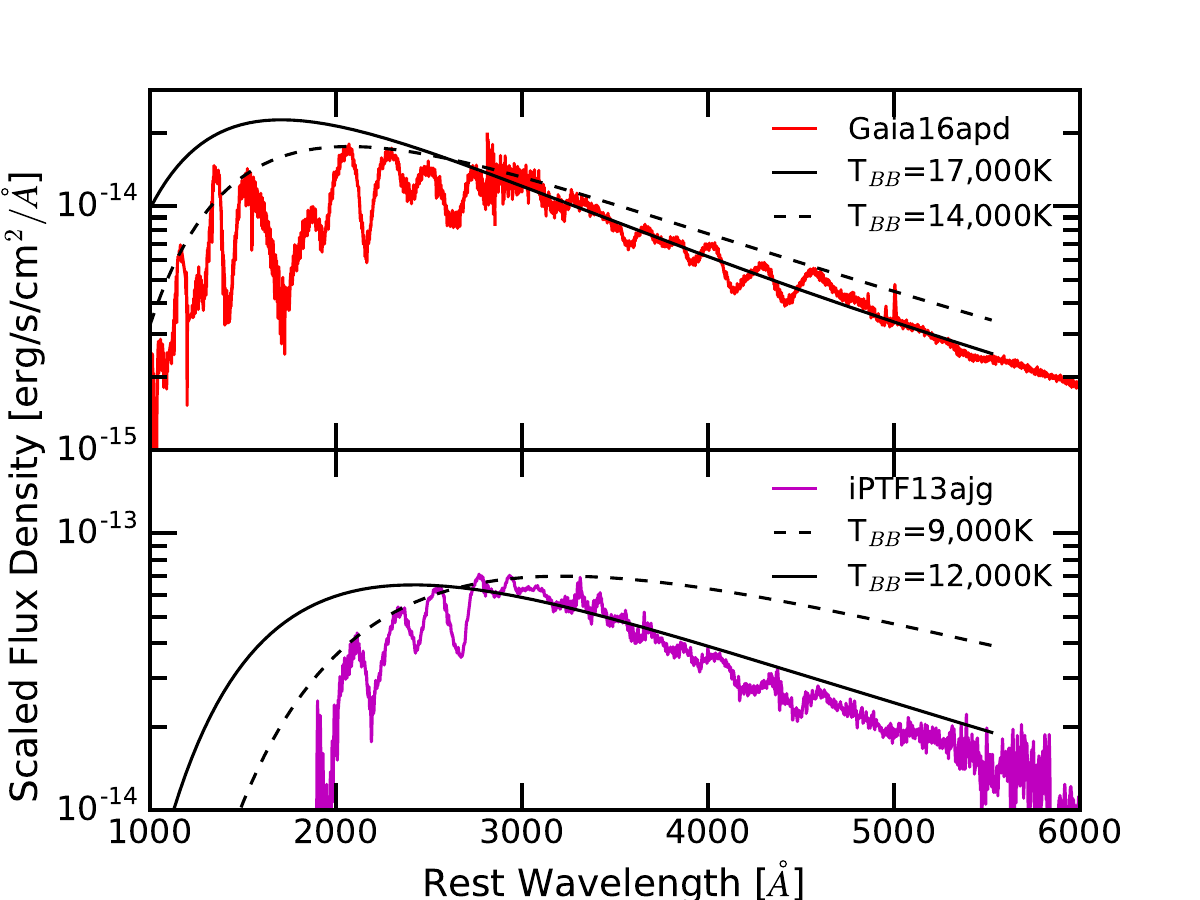}
\caption{ The plot shows that a pure blackbody function can not fit the observed spectra. \label{fig:black}}
\end{figure}

Figure~\ref{fig:uvspec} presents the {\it HST} spectrum of SN2017egm, in comparison with those of Gaia16apd, PTF12dam and iPTF13ajg. The spectra were corrected for the Galactic extinctions (Table~\ref{tab:date}) and arbitrarily scaled for visual clarity. The two example spectra in Figure~\ref{fig:black} illustrate that the SLSN-I UV and optical continuum cannot be simultaneously fit by a blackbody with a single temperature. {\color{black} This finding was also noted by \citealt{Chomiuk2011} for SLSN-I PS1-10ky at $z=0.9$.}  These two figures together demonstrate that the difference in UV SED shape between Gaia16apd and iPTF13ajg can not be explained {\color{black} by the difference in blackbody temperature alone.}

\begin{figure}[t!]
\includegraphics[width=0.5\textwidth]{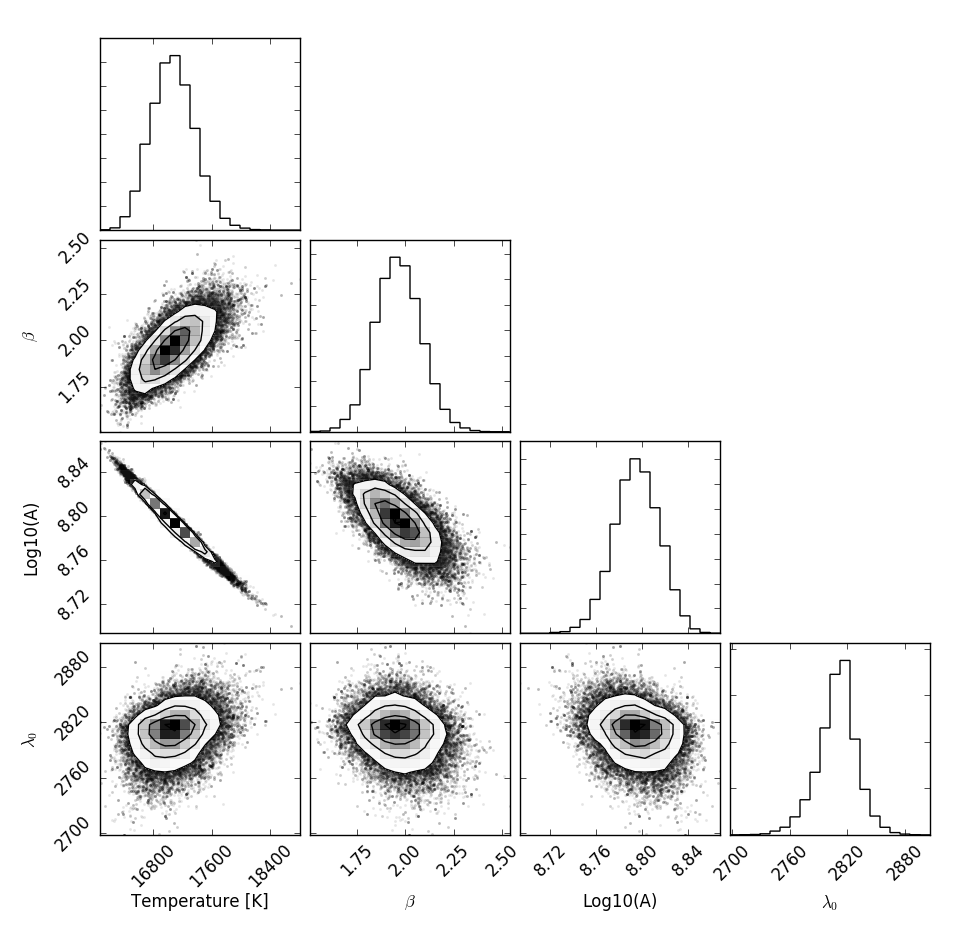}
\caption{ The probability distributions of the four variables ($T$, $\beta$, $\lambda_0$ and $\log_{10}(A)$ derived from the MCMC simulations for SN2017egm.   \label{fig:mcmc}}
\end{figure}

{\color{black} To characterize UV \&\ optical spectral continuum, we use Markov Chain Monte Carlo (MCMC) method, specifically utilizing python emcee package \citep{Foreman-Mackey2012}. Our adopted model is a modified blackbody, defined as 
as $f_{\lambda}(T) = (\lambda/\lambda_0)^\beta f_{\lambda,BB}(T)$ for $\lambda < \lambda_0$ and $f_{\lambda}(T) = f_{\lambda,BB}(T)$ for $\lambda > \lambda_0$.  We do not use all of the pixels from the observed spectra and exclude the spectral regions with prominent broad absorption features. We resample the spectra by selecting a number of spectral regions (with width of $50-150$\AA) presenting the continuum, and compute the averaged flux density from each spectral bin. 
For example, the selected regions used for the MCMC calculations are shown as solid black points for SN2017egm in Figure~\ref{fig:uvspec}, with each error bars in X-axis marking the selected regions.

Our model has four variables, 
$T$, $\lambda_0$, $\beta$ and $\log_{10}(A)$ (a scaling factor taking into account of distance).  Table~\ref{mcmcoutput}  shows the results from the computation where all four parameters are considered as variables.  Figure~\ref{fig:mcmc} shows the probability distributions of these four variables calculated from the MCMC simulation for SN2017egm.  We find that $\lambda_0$ values are similar, $\sim2800$\AA, for the three SLSN-I, except Gaia16apd. As Table~\ref{mcmcoutput} shown, Gaia16apd has $\lambda_0$ at a longer wavelength.  The derived UV power-law slope $\beta$ is small, 0.51.  This is not surprising because $\lambda_0$ and $\beta$ have some degree of inter-dependency, larger $\lambda_0$ value would lead to smaller $\beta$. Therefore, to make a meaningful comparison of $\beta$ values, it makes more sense to fix $\lambda_0$ at $2800$\AA\ and recompute the other three parameters.  Table~\ref{mcmc2} shows the MCMC simulation results when considering only three variables, $T$, $\beta$ and $\log_{10}(A)$. In the remaining of the paper and the relevant figures, we quote the $T$, $\beta$ values derived from the MCMC calculations with the fixed $\lambda_0$. 

The spectral continua at $\lambda>2800$\AA\ can be well described by a blackbody with $T_{BB} \sim 17094^{+587}_{-285}$\,K for SN2017egm, $16703^{+480}_{-398}$\,K for Gaia16apd, and slightly cooler temperature of $15008^{+200}_{-75}$\,K for iPTF13ajg  and $14684^{+87}_{-76}$\,K for PTF12dam.  This value is smaller for iPTF13ajg than derived by \citet{Vreeswijk2014}. The lower temperature for PTF12dam is consistent with the phase when the spectra were taken, when the ejecta has a longer time to cool.

At $\lambda < 2800$\AA, SLSN-I UV continuum is significantly below the blackbody model. The UV deviation from a blackbody emission (or UV continuum suppression) is well known for normal supernovae, and is generally thought due to a forest of line absorption by heavy elements (so called line blanketing at $\lambda < 1300$\AA) as well as UV photons scattered to longer wavelengths by the fast expanding ejecta \citep{Pauldrach1996,Bufano2009}.  It is also known that at maximum light, SLSNe-I tend to be bluer and the UV continuum suppression is much less than that of normal SNe \citep{Quimby2011, Yan2017}.

Comparison of Table~\ref{mcmc2} suggests a robust conclusion that SN2017egm has $\beta = 0.8^{+0.15}_{-0.16}$, much smaller than $\beta =  2.94^{+0.09}_{-0.04}$ for iPTF13ajg.}
Another way of describing the UV continuum variation is  the 1400\AA\ to 2800\AA\ continuum flux ratio, which changes from 1.5 in Gaia16apd, to 0.5 in iPTF13ajg. \citet{Nicholl2017b} also uses a modified blackbody SED with $\beta = 1$ to model a sample of SLSNe-I. Our analysis shows that the UV SED variation is large enough that a single SED template is not sufficient when modeling samples of SLSNe-I. The large near-UV SED variation is also noted in \citet{Lunnan2017}, which found that for a sample of SLSNe-I at $z\sim 0.5 - 1.5$, their peak absolute magnitudes at near-UV (2600\AA) span over 4 magnitudes, whereas the peak magnitudes at 4000\AA\ spread over only 2 magnitudes.

\begin{figure}
\includegraphics[width=0.5\textwidth]{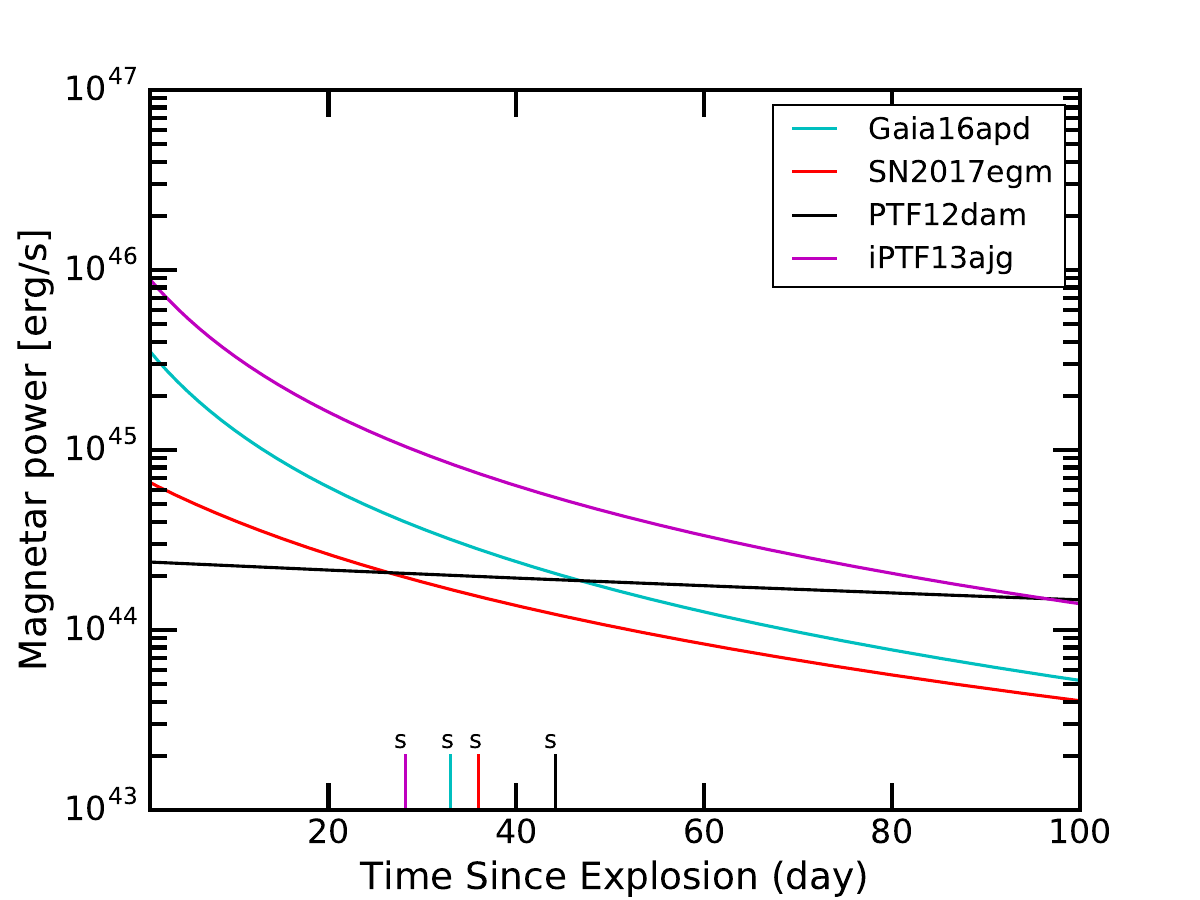}
\caption{ The magnetar power for each event in our sample. The magnetar model parameters for these four events are taken from \citet{Nicholl2017c}. \label{fig:power}}
\end{figure}

The physical origin of the UV variation in SLSNe-I is not clearly understood.
\citet{Nicholl2017c} has argued that the extremely high UV continuum at maximum light from Gaia16apd could be explained by its high magnetar power and the small photospheric radius. In Figure~\ref{fig:power}, we plot the magnetar power as a function of time since explosion, 
described by the equations:
$$t_{spin} = 1.3\times10^{5}(\frac{M_{NS}}{1.4M_\odot})^{3/2}(\frac{B}{10^{14}G})^{-2}(\frac{P}{1ms})^{2}\ \ {\rm s}$$

$$F_{mag} = 2\times10^{47}(\frac{B}{10^{14}G})^2(\frac{P}{1ms})^{-4}(1.0+\frac{t}{t_{spin}})^{-2}\ \ {\rm erg/s}$$

We choose the time since explosion because it is directly related to ejecta cooling, thus the appearance of the observed spectra. 
We adopt the magnetar model parameters for these four events from \citet[][Appendix A]{Nicholl2017b}, derived using {\it MOSFiT} \citep{Guillochon2017}. At the time of the spectral observations (vertical line in Figure~\ref{fig:power}), iPTF13ajg has the most powerful heating source $F_{mag}$, but its far-UV continuum is the most suppressed. {\color{black} This implies that magnetar power is not directly proportional to UV continuum emission.  \citet{Nicholl2017b} noted that photospheric radius $R$ may play an important role, in addition to the magnetar power.  
However, we note that the regions producing far-UV continuum photons might not be the same as the photospheric surface, which is usually determined by assuming color temperature $T_{BB}$ equals to effective temperature $T_{eff}$, defined as $L_{bol} = 4\pi\sigma R^2 T_{eff}^4$ (see discussion below). The significant deviation from a single blackbody fit at the UV wavelength appears to support the idea that UV photons may come from different regions than that of optical photons.  Therefore, we need a better understanding of the radiative physics of early-time SLSN-I spectra.}



\begin{figure}[h!]
\includegraphics[width=0.55\textwidth]{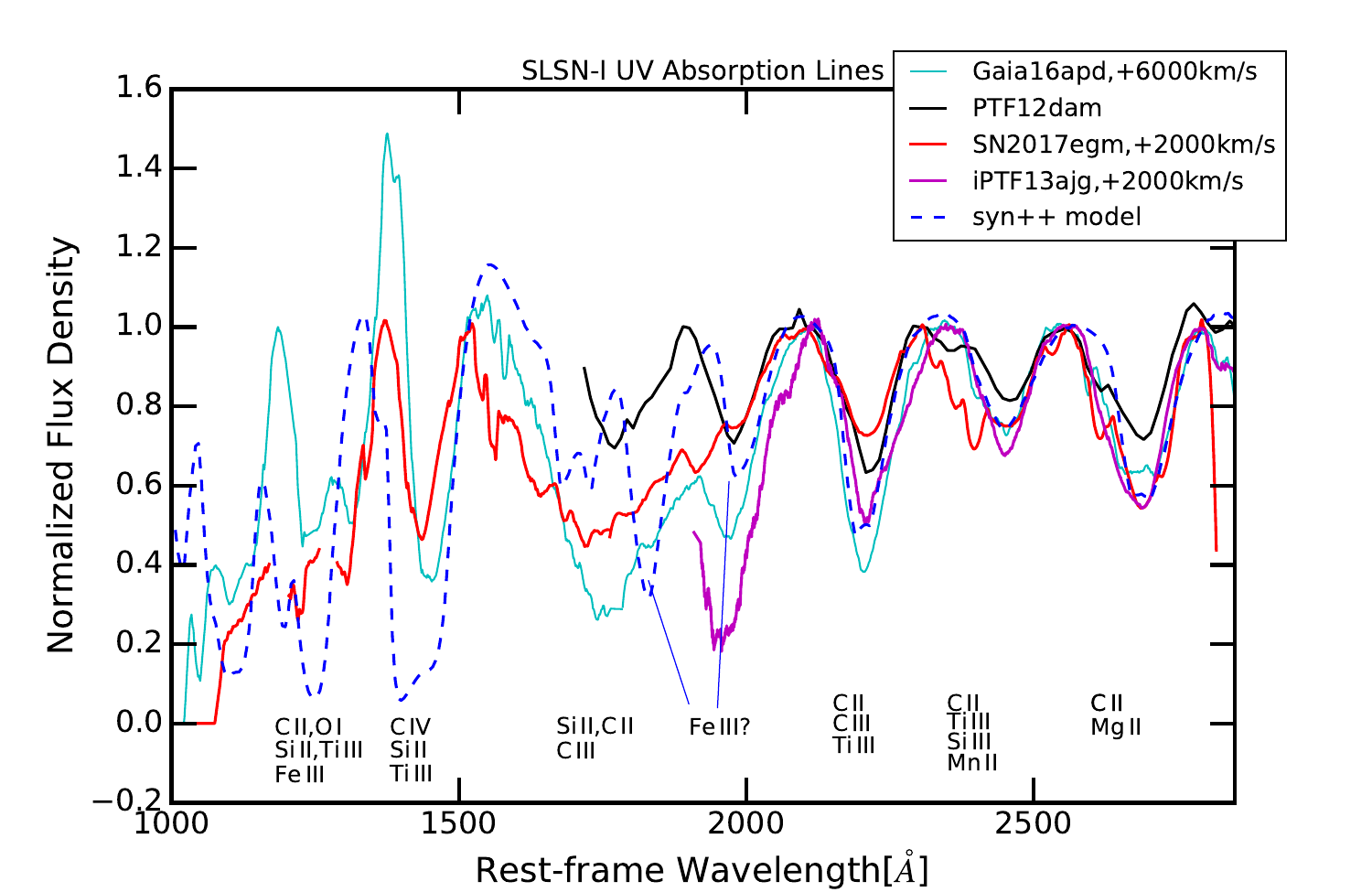}
\includegraphics[width=0.55\textwidth]{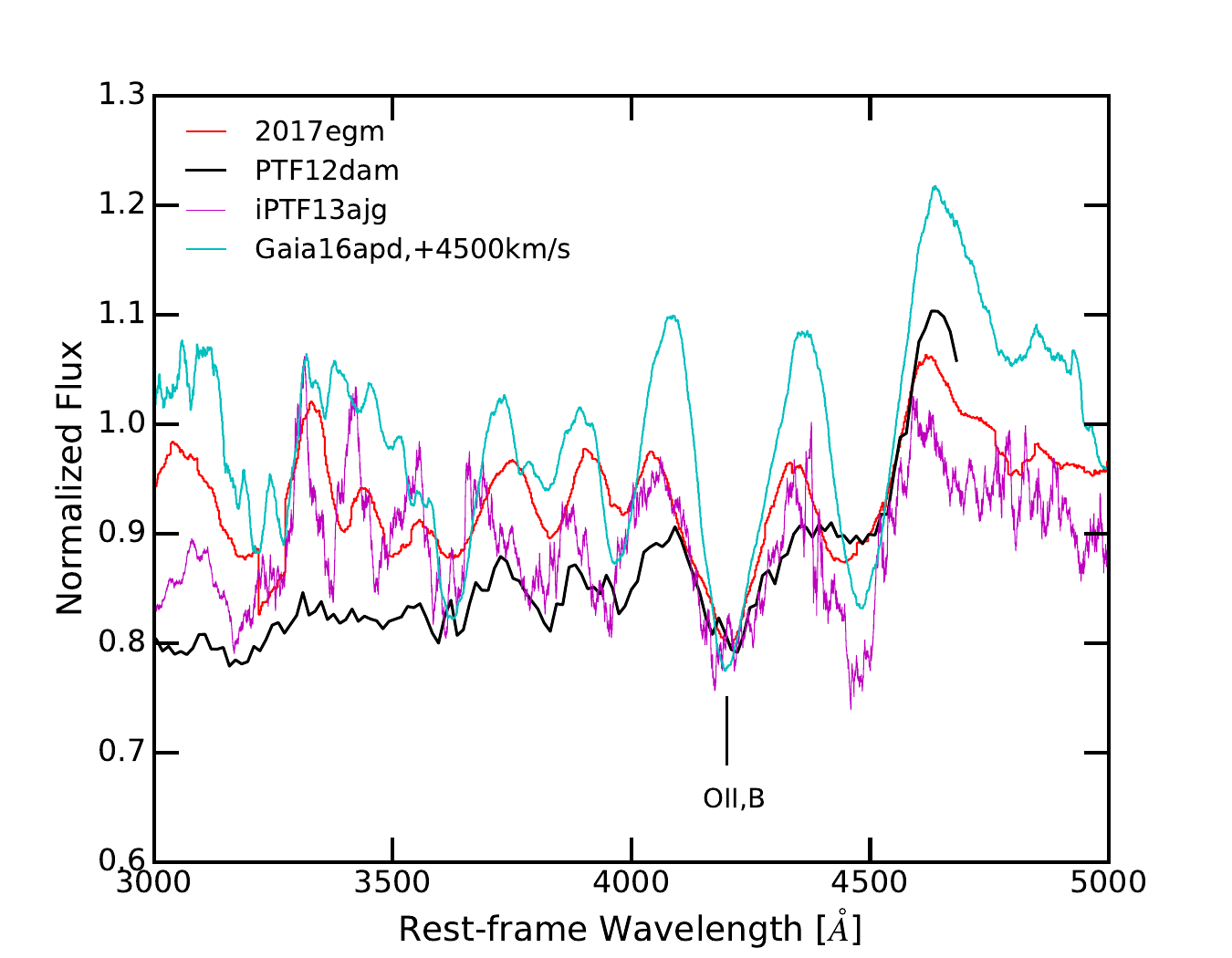}
\caption{ The normalized spectra in comparison with the model spectrum (blue dashed line) generated by Quimby et al. (2017) using syn++ code \citep{Thomas2011}. {\color{black} To align the spectral features, we redshifted the spectra by the indicated velocities. The shifts are generally different for UV and optical spectra for the same SLSN-I.}  \label{fig:norm}}
\end{figure}

Quimby et al. (2017) carried out a detailed analysis of near-UV spectra of 8 SLSNe-I. They identified three common features at 2200, 2450 and 2650\AA, likely produced by C\,II, C\,III, Ti\,III, Mg\,II and Mn\,II \citep[Also see][]{Howell2013,Mazzali2016,Dessart2012}. Here we focus on the far-UV, and plot out the normalized spectra in Figure~\ref{fig:norm}, where the top panel shows the far-UV spectra from 1000 - 3000\AA, and the bottom panel shows the optical portion between 3000 - 5000\AA. {\color{black} To align the spectral features, we redshifted the spectra by the respective amount indicated in the figure. The shifts are different for UV and optical spectra of the same object.} We used the local continuum to normalize each spectrum. 
Because velocities for PTF12dam are well measured in Quimby et al. (2017), we shift other spectra relative to PTF12dam in order to get estimates of the velocities. 
In the optical, we align all spectra by O\,II absorption at 4200\AA\ (feature B, Quimby et al. 2017), and in the UV, we align the spectra by the three near-UV features. Figure~\ref{fig:norm} also includes the synthetic spectrum (blue dashed line) generated by syn++ model \citep{Thomas2011} and presented in Quimby et al. (2017).  

With the first epoch {\it HST} spectrum at $\Delta t_{exp}=+44.2$\,day, PTF12dam has an O\,II velocity of $\sim$10,000\,km\,s$^{-1}$. From the comparison, SN2017egm and iPTF13ajg have the same O\,II velocities, but at earlier phases ($\Delta t_{exp}$\,=\,36 \&\ 28\,d respectively). Gaia16apd is different, has a much higher velocity, $\sim$14,500\,km\,s$^{-1}$ at the maximum light ($\Delta t_{exp}$\, =\,33\,d), consistent with what we derived in \citet{Yan2017}. In addition, the UV spectral comparison appears to give different results. Gaia16apd, SN2017egm and iPTF13ajg are +6,000, +2,000, +2,000\,km\,s$^{-1}$ faster than PTF12dam respectively. This implies that far-UV spectral formation layers are moving faster than the O\,II ions which produce the absorption series at 3500 - 4700\AA. 

In the published literature on SLSNe-I, it is commonly assumed that blackbody temperature $T_{BB}$ (also called color temperature) is the same as photospheric effective temperature (not directly observed) $T_{eff}$, defined as $L_{bol} = 4\pi\sigma R^2 T_{eff}^4$. Using this equation, we can infer photospheric radius $R$ as well as photospheric velocity $v_{phot} = R/\Delta t_{exp}$, here $\Delta t_{exp}$ is the time since explosion. We find the ratio $v_{phot}/v_{OII}$ of 0.49, 0.51, 1 and 0.62 for Gaia16apd, SN2017egm, iPTF13ajg and PTF12dam respectively. Only iPTF13ajg appears to have O\,II velocity close to its photospheric velocity, and other three events with less UV suppression seem to have O\,II ions expanding faster than photosphere. One might argue the small photospheric velocities could be due to uncertainties in $\Delta t_{exp}$.  It seems unlikely that for all three events $\Delta t_{exp}$ values are systematically over-estimated by a factor of 2, which would make the true rise times for Gaia16apd and SN2017egm $\sim$15\,days. This is too short for a SLSN-I {\color{black} \citep{Nicholl2015, DeCia2017b}}, making this explanation unlikely.  Another possibility is that the assumption of $T_{BB} = T_{eff}$ is no longer valid, with $T_{eff} \le T_{BB}$, which would imply larger photospheric radius $R$ and $v_{phot}$. 
As discussed in \citet{Nakar2010}, $T_{eff}$ in $L_{bol} = 4\pi\sigma R^2 T_{eff}^4$ is based on the assumption of thermal equilibrium. The observed temperature $T_{BB}$ is mostly determined by the outer layers of ejecta. If these layers are not in thermal equilibrium, the observed temperature $T_{BB}$ would be $>T_{eff}$. We conclude that the high velocity gas layers which make significant contributions to the observed spectra for Gaia16apd, SN2017egm and PTF12dam are likely not in thermal equilibrium.

The far-UV {\it HST} spectra show three new absorption features at 1250, 1400 and 1650\AA, detected in both SN2017egm and Gaia16apd (also partially in PTF12dam), and with similar equivalent widths. We note that the 1650\AA\ feature is particularly broad and strong.  We use the synthetic spectrum to make a tentative spectral identifications. The mock spectrum includes ions such as Fe\,III, Ti\,III, Ti\,II, Si\,II, Mg\,II, O\,II, O\,I, C\,IV, C\,III and C\,II, and the detailed discussion can be found in Quimby et al. (2017). We identified the 1250\AA\ feature as the blend of C\,II, O\,I, Si\,II, Ti\,III and Fe\,III, and the 1400\AA\ feature from C\,IV, Si\,II and Ti\,III.  However, the most prominent absorption at 1650\AA\ is poorly explained by the synthetic spectrum. This feature blends into the absorption at 1950\AA\ which was identified as Fe\,III, and possibly also Co\,III \citep{Mazzali2016}. 
{\color{black} It is possible that the broad feature at 1650\AA\ is a blend of Si\,II, C\,II, C\,III and Fe\,III, with strength and broadening significantly different from the syn++ calculated spectrum. Better line identifications at the far-UV are needed in future studies.}
 
In summary, the seven broad UV absorption features are stronger than the features in the optical regime in the early-time spectra of SLSNe-I. We find that their strengths, {\it i.e.} equivalent widths, appear to be remarkably similar within a factor of two among these four events. This was also noted in \citet{Nicholl2017c} for Gaia16apd and iPTF13ajg.  
The uniformity and large equivalent width of UV spectral absorption features bode well for high-$z$ SLSN-I searches since ground-based follow-up spectroscopy will rely on them for solid classifications.

\subsection{Interstellar and circumgalactic media along the line-of-sight to SN2017egm} 
 
The far-UV spectrum of SN2017egm provides a sensitive probe of ISM absorbers at different velocities along the line-of-sight and enables the measurements of metal abundances, thus shedding light on the SN host environment.

\begin{figure*}[h!]
\includegraphics[width=0.99\textwidth]{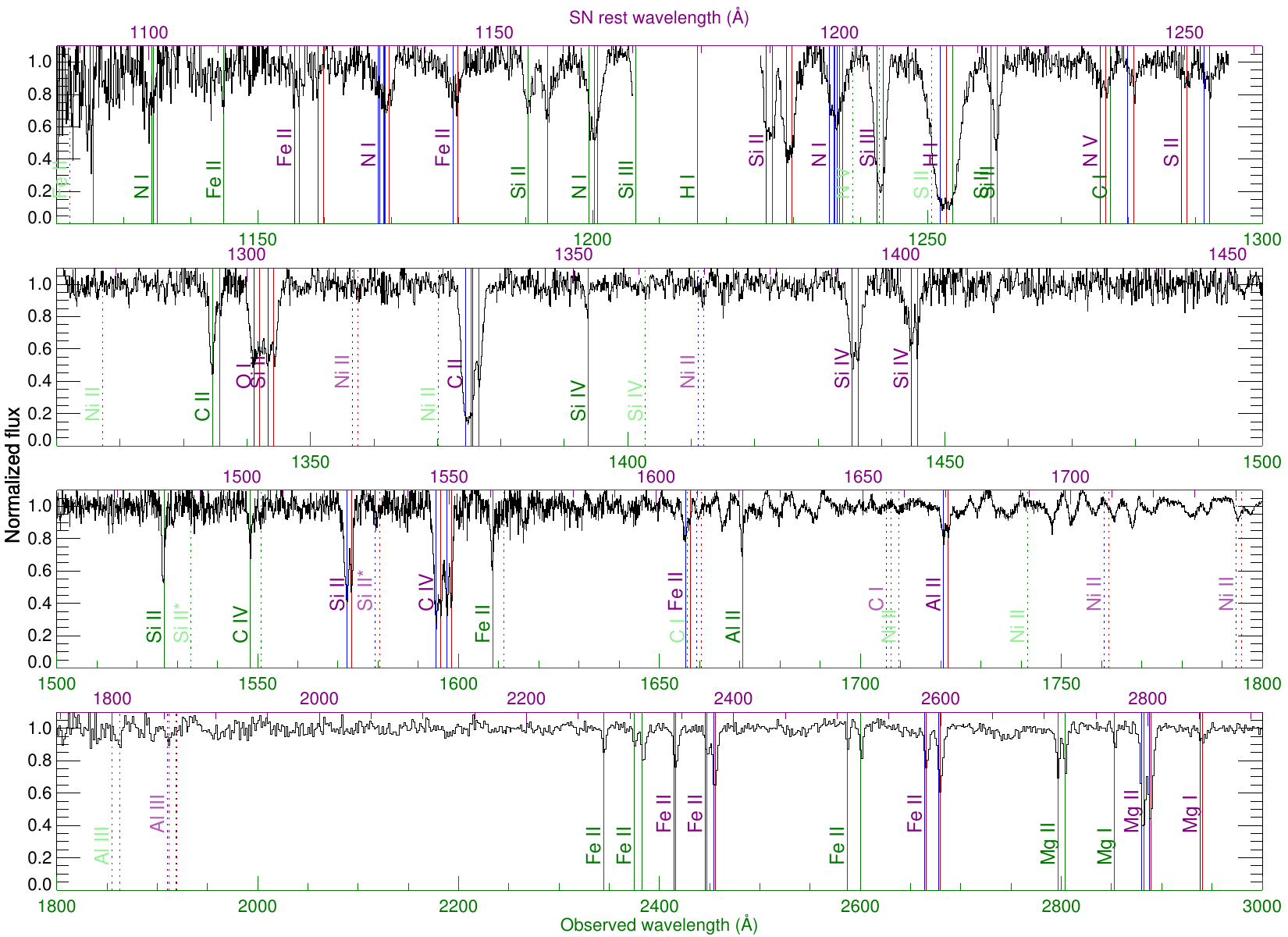}
\caption{\emph{HST} COS and STIS spectra of SN\,2017egm after normalization of
the continuum using Savitsky-Golay convolution. Absorption
lines originating from the Milky Way Galaxy are marked as green, while
lines from the host system are marked as red (host component) or blue
(companion component) and jointly labeled with purple text.  Dotted
lines and lighter text colours indicate notable nondetections or
marginal detections.  The resolution in the mid-UV (bottom-most panel)
is not sufficient to resolve the two velocity components in the host
system. \label{fig:fullspec}}
\end{figure*}

We first applied a Savitsky-Golay filter to the SN spectrum to produce a model of the complex continuum emission, including the broad SN features. Strong narrow lines were interpolated over prior to applying the filter. We then divided the observed spectrum by the model spectrum to produce a normalized spectrum, shown in Figure~\ref{fig:fullspec}.  This spectrum clearly reveals several sets of absorption systems: one at $z=0$ from the Milky Way Galaxy, and two more related to the host system.  The first, at 
$z$\,=\,$0.0307$, is effectively consistent with the emission-line redshift of NGC\,3191.  The other system is at $z=0.0299$, corresponding to a relative blue-shift of 235\,km\,s$^{-1}$.  

Although there is no other galaxy directly along the line of sight towards NGC\,3191, at 73$\arcsec$ to the west (44\,kpc in projection) a companion galaxy, SDSSJ101857.98+462714.6, is visible.  The spectroscopic redshift of this system is $z=0.02993$: consistent with our second absorption system, making this galaxy a plausible counterpart of that system.  
Alternatively, the blue-shifted system could be due to an outflow or wind emanating from NGC\,3191, as has been seen in some other high-SFR low-$z$ galaxies \citep{Grimes2009,Heckman2015,Alexandroff2015}.  

Our spectroscopic coverage includes the damped Lyman-$\alpha$ absorption (DLA) feature from neutral hydrogen.  Analysis of this feature is complicated by blending between the host and the companion, as well as uncertainties in the continuum modeling, and it is difficult to constrain the individual contributions of the two component systems to the overall line. Modeling the line as two separate, blended Voigt profiles, we can robustly constrain the total column density of the two systems together to $\rm 5\times 10^{19} cm^{-2} \le N(HI) \le 9\times10^{19} cm^{-2}$; our best-fit estimate is $(6.7\pm1.2)\times$ $10^{19}$\,cm$^{-2}$, making the system technically a sub-DLA \citep{Peroux2003}. This spread in N(HI) is due to uncertainties in the relative contributions of the two systems and continuum placement. The derived N(HI) is an order of magnitude less than a typical value through a disk of a spiral galaxy with stellar mass $>10^{10}M_\odot$, measured by The HI Nearby Galaxy Survey (THINGS) \citep{Walter2008, Leroy2008}. 
This result has two implications. One is that SN2017egm may have exploded on the near side of the galaxy, where there is less neutral HI material along the line-of-sight. One may argue for the second possibility that UV fluxes from the SN explosion could photo-ionized a large fraction of neutral HI in the disk of NGC\,3191, and SN2017egm could be anywhere in the disk. 
To validate this hypothesis, we calculate the time required, $t_{phot}$, to photo-ionize N(HI) $\sim10^{21}$\,cm$^{-2}$ over a scale height $R$ of an HI disk. HI-ionizing flux, $J_{UV}$ at the time of SN explosion is poorly constraint by observation. Let us take $J_{UV}$ as a fraction of $f_{UV}$ of the peak bolometric luminosity ($L_{bol}$),  we have $J_{UV} = \frac{f_{UV} L_{bol}}{4\pi R^2 (\rm 13.6eV)}$. {\color{black} The Milky Way thin disk (stellar) has a scale height of 100\,pc \footnote{the HI gas can be more extended \citep{Marasco2011}}. As the most conservative assumption, let us take $R = 50$\,pc if the SN is at the mid-plane}, we have $J_{UV} =3\times10^{11}(f_{UV}/0.01)$\,cm$^{-2}$\,s$^{-1}$. {\color{black} In this equation, without any knowledge of the early-time UV fluxes from SN\,2017egm,  we assume it is only 10\%\ of the estimated bolometric luminosity ($f_{UV} = 0.01$).} Here the maximum $L_{bol}$ is $2\times10^{44}$\,erg\,s$^{-1}$. To photo-ionize a column of $10^{21}$\,cm$^{-2}$ HI atoms, we need the same number of HI ionizing photons, thus, over a time scale of $t_{phot}$, we have $J_{UV} t_{phot} \simeq 10^{21}$\,cm$^{-2}$, leading to $t_{phot} = 3\times10^9$\,second. This is two orders of magnitude longer than the time lag between the explosion and the {\it HST} spectroscopy date for SN2017egm. {\color{black} Since $t_{phot} \propto R^2 f_{UV}^{-1}$, a larger $f_{UV}$ value will shorten the photoionization time scale $t_{phot}$ by a factor of a few, but not sufficient to change our conclusion.}  This result suggests that photo-ionization due to the SN explosion is probably localized within a 5\,pc region.

\begin{figure}[h!]
\includegraphics[width=0.4\textwidth]{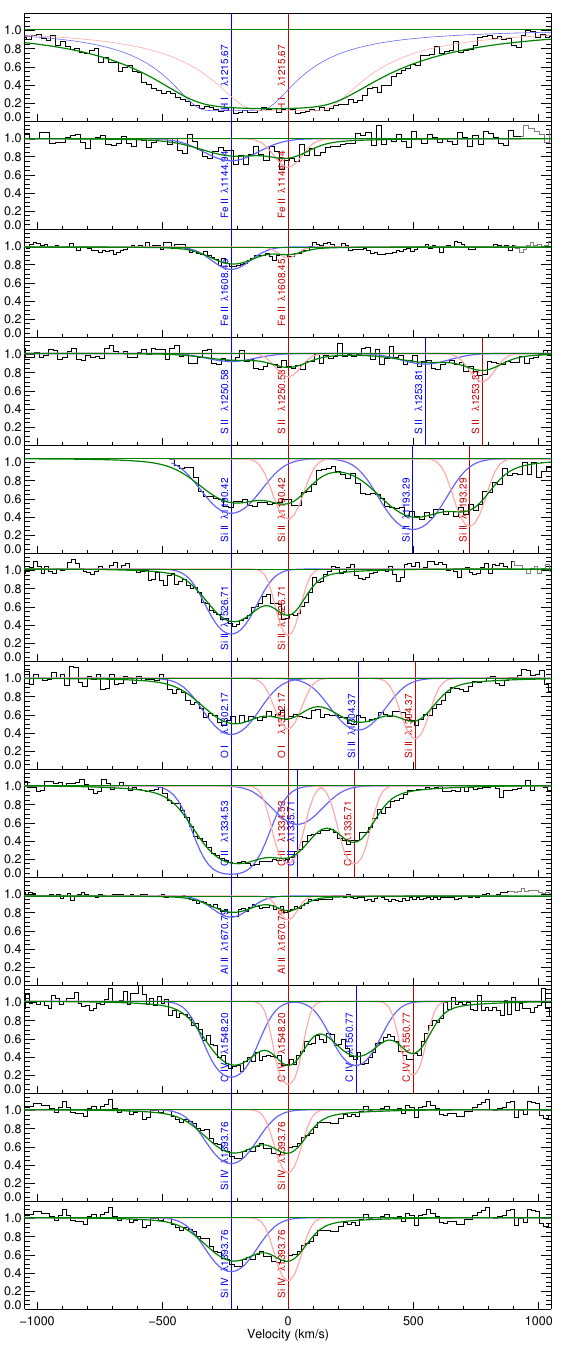}
\caption{Profiles of prominent absorption lines from various species associated with the host-galaxy system of SN 2017egm (see also Figure~\ref{fig:fullspec}).  All lines except for the sub-DLA (top plot) show a bimodal profile, with the red component at a redshift matching that of NGC\,3191 and the bluer component (-235\,km\,s$^{-1}$) at the exact redshift of the companion galaxy SDSSJ101857.98+462714.6.  Red and blue curves show modeled intrinsic profiles and the green curves are the combined results after convolution with the COS instrumental line-spread-function (LSF).  
\label{fig:linefit}}
\end{figure}

Using the calibrated relation of $\rm N(HI) = 6.86\times10^{21}\times E(B-V)$ \citep{Guever2009}, we derive $\rm E(B-V)$, the host dust extinction of $0.007$. This relation is based on Galactic dust-to-gas and dust-to-metal ratios and should work reasonable well for galaxies with high metallicities.
The $\rm E(B-V)$\,=\,$0.3$ \citep{Chen2017b,Izzo2017}, derived from nebular emission lines at the SN location, represents the dust extinction through the full disk of {\color{black} NGC\,3191}. The low host extinction correction is further supported by the fact that the blackbody temperature derived from the uncorrected SN2017egm spectra appears to be consistent with that of other SLSNe-I in dwarf galaxies. 

Detailed abundance analysis of the metal lines is complicated by the relatively low resolution (FWHM\,$\sim$\,0.47\AA\,$\sim$\,120\,km\,s$^{-1}$) of the COS G140L grating, which means that most lines strong enough to detect are also saturated. Additional continuum normalization uncertainties and S/N limitations impose other uncertainties. {\color{black} If a line is saturated, its equivalent width, {\it i.e.} line strength, becomes less sensitive to column density of the corresponding ion, making abundance measurement unreliable. Si\,II\,1190, 1193, 1526\AA\ lines are strong lines and commonly used for metallicity measurements, however, they are saturated in our data. This is validated by checking their equivalent widths, which are not proportional to their oscillator strengths for the same ion column density.}


Three weaker lines, S\,II\,$\lambda\lambda$1250,1253\,\AA, Fe\,II$\lambda$1608\AA\ and Al\,II$\lambda$1670\AA\ are not saturated and offer a more promising route to constrain the abundances. They are weak but still significant detected. 
In the optically thin regime, the column density N(X) of an ion X is directly proportion to line strength, with N(X)\,=\,$\tau [m_e c^2/(\pi e^2 f \lambda^2)]$, {\color{black} and $\tau$ is the integrated optical depth over a whole line, $f$ is the oscillator strength, $m_e$, $e$ and $c$ are electron mass, electron charge and speed of light respectively.  $\tau$ is computed from the following equation, 
$\tau(\lambda) = [A/(\sqrt{2\pi}b_D)]\exp[-(\lambda-\lambda_0)^2/(2b_D^2)]$, with $A$  the integrated line flux, $b_D$ is the line width and the derived line profile is $flux(\lambda) = \exp[-\tau(\lambda)]$.  As illustrated by the equation, the optical depth profile is a Gaussian and the line profile is an exponential. The observed line profile is the convolution product of the exponential profile with the instrument Line-Spread-Function (LSF). Line width $b_D$ is intrinsic line width. We explored various fitting parameters and find that the $b_D$ values of 0.4 and 0.2\AA\ achieve reasonably good fits to all of the absorption lines.  }

Table~\ref{tab:line} listed the column densities and abundances for the metal lines at these two velocities.  {\color{black} Figure~\ref{fig:linefit} shows the fits to various metal lines and broad Lyman-$\alpha$ absorption. The best fit to Lyman-$\alpha$ requires two almost equal strength absorption components with a velocity separation of 235\,km\,s$^{-1}$.  The $\chi^2$ values from the fits suggest that the probability of Lyman-$\alpha$ being a single line at one redshift is very small, $<2$\%.
The derived HI column densities are $(3.0\pm0.8)\times10^{19}$ and $(3.7\pm0.9)\times10^{19}$\,cm$^{-2}$ for the two absorbers at $z=0.0307$ (redshift of NGC3191) and $0.0299$ (companion galaxy) respectively.
Element abundance [X/H] is calculated as [X/H] = 12 + $\log_{10}{(\rm N(X)/N(HI)) - [X/H]_\odot}$, and the corresponding error considers both N(X) and N(HI) errors quadratically. In Table~\ref{tab:line}, [X/H]$_{\rm obs}$ 
indicates the abundance values computed directly from the column densities without any other corrections. We find [S/H]$_{obs}$ of 0.30 and 0.38 (2.0 \&\ 2.4~$Z_\odot$) for the two velocity component respectively.}

There are several effects to consider. One is ionization correction which accounts for the fact that some H and S atoms are ionized \footnote{H and S ionization potentials are 13.6 and 23\,eV respectively.}. Ionization correction for H is adopted from a {\it HST} UV spectroscopy study of a sample nearby sub-DLAs by \citet{Werk2014}. We adopt H ionization fraction of 72\%\ at the corresponding HI column density of $6.7\times10^{19}$\,cm$^{-2}$ \citep[][Table 1]{Werk2014}. Ionization correction for S is much more difficult and is not considered here. Even without this correction, the [X/H]$_{\rm HC}$ including H ionization correction should serve meaningful constraints on the lower limits to the abundances. We conclude that [S/H] abundance should {\color{black} be greater than [S/H]$_{\rm HC}\sim$0.06 and 0.14 for the two velocity components respectively (1.15 and 1.4~$Z_\odot$, Table~\ref{tab:line}).}

Another effect to consider is dust depletion of S, Fe and Al, which reduces the strengths of the absorption lines. This is why the observed [Fe/H]$_{\rm obs}$ values are very low, $-0.62$ and $-1.36$. {\color{black} The large observed [S/Fe]$_{obs}$ and [S/Fe]$_{obs}$ unambiguously imply a substantial amount of dust, comparable to that of the Milky Way. Although dust depletion correction to metal elements are uncertain,  we proceed and adopt the method based on [S/Fe] ratio and described in \citet{deCia2016,deCia2017}.}  We derived the dust depletion corrected $\rm [Fe/H]_{DC}$ ratios of 0.48 and 1.10 (3 and 12~$Z_\odot$). Including both the dust depletion and H ionization corrections, the final [Fe/H]$_{\rm corr}$ values are {\color{black} 0.25 \&\ 0.87 (1.8 and 7.4~$Z_\odot$)} for the two velocity components. 

{\color{black} After both H ionization and dust depletion corrections, S, Fe and Al abundances at $z=0.03070$ and $0.0299$ are all super-solar, except [Al/H] at $z=0.0307$, which is $-0.25$ (0.56~$Z_\odot$).  All six metal abundance values indicate a wide range of metallicities for the two absorbers. Such a wide range reflects the difficulties of this type of measurements using absorption lines from a moderate resolution spectrum.  If we take the average value of the derived abundances, the result is clearly above solar metallicity. Therefore, taking together the six metal abundance measurements, we conclude that the environment of SN2017egm likely has a solar or higher metallicity.}
For comparison, the median metallicity of 40 SLSN-I host galaxies published in the literature is $0.3 Z_\odot$ on the PP04 N2 scale \citep{Pettini2004, Chen2017b}. Both nebular emission line and UV absorption line analyses indicate that the metallicities at the SN location and the companion galaxy are at least a factor of 3 higher than that of the host galaxies of most SLSNe-I at low redshift.

Considering the low HI column density alone, one might argue that there could be a dwarf companion galaxy at the location of SN2017egm. However, the high metallicity result makes this scenario very unlikely since dwarf galaxies tend to be metal poor.

\section{Summary}


We present the maximum light {\it HST} UV \&\ optical spectra of the closest SLSN-I, a rare event exploded in a massive spiral galaxy. Using this data, together with three other SLSNe-I, we address the question of how early-time UV continuum and spectral features of SLSN-I vary from source to source. In addition, we set constraints on the metal abundances of local environment of SN2017egm by performing a detailed analysis of narrow UV absorption lines detected in the {\it HST} spectra. 

High quality early-time UV spectra of four SLSNe-I appear to detect a common set of broad UV absorption features (seven) between $1000 - 2800$\AA. And the three features at the far-UV $1000 - 1500$\AA\ are new, revealed only by the {\it HST} data for SN2017egm and Gaia16apd \citep{Yan2017}. These UV features appear to be much stronger than ones in the optical regime at similar phases. And furthermore, their absorption strengths characterized by their equivalent widths are similar within a factor of 2 from source to source.  {\color{black} The physical implication of this result is worth further exploring in the future.  It should provide data constraints on spectral formation models for SLSN-I, which will help us better understand the characteristics of outer photospheres of SLSN-I at pre-peak phases.} These results {\color{black} also} bode well for optical spectral classification of SLSN-I candidates at $z>2$.   

In contrast to the uniformity of the UV absorption features, the $1000-3000$ \AA\ continua vary significantly. 
{\color{black} We quantify the UV continuum variation by using a modified blackbody function, 
$f({\rm T_{BB}},\lambda) = {\rm B(T_{BB}},\lambda) (\lambda/\lambda_0)^\beta$ for $\lambda < \lambda_0$, and $f(T_{BB},\lambda) = B(T_{BB},\lambda)$ for $\lambda > \lambda_0$.  If considering $T$, $\beta$ and $\lambda_0$ all as the variables, our MCMC calculations find that SN2017egm, iPTF13ajg and PTF12dam all have $\lambda_0 \sim 2800$\AA.  Since $\beta$ and $\lambda_0$ have some degrees of inter-dependency, in order to compare the far-UV continuum slopes, we fix $\lambda_0$ at $2800$\AA\ for all four SLSN-I.  The MCMC calculations find that the parameter $\beta$, describing the decline of the far-UV continuum toward the shorter wavelength, is $0.80^{+0.15}_{-0.16}$, $1.96^{+0.15}_{-0.14}$, $1.78^{+0.03}_{-0.03}$ and $2.94^{+0.09}_{-0.04}$ for  Gaia16apd, SN2017egm, PTF12dam and iPTF13ajg respectively. Our SED model is useful for calculating bolometric corrections and modeling of light curves using only optical photometry. }  Using PTF12dam as the reference, we find Gaia16apd has the highest velocity, 14,500 - 16,000\,km\,s$^{-1}$, and the ionized gas producing UV spectral features are clearly moving faster than that of O$^+$ ions forming O\,II absorption at 3500 - 4700 \AA.   
  
We find the observed parameters such $T_{BB}$ and $v_{OII}$ can not be consistently related to the photospheric temperature and velocity $T_{eff}$ and $v_{phot}$ where $T_{eff}$\,=\,$(L_{bol}/4\pi\sigma (v_{phot} \Delta t_{exp})^2)^{1/4}$. This suggests that the ejecta layers producing continuum emission (photosphere) and absorption features are probably not in thermal equilibrium and their expansion speeds are likely different.  


The {\it HST} SN2017egm spectra also detect a rich set of narrow absorption lines produced by gas absorbers along the line-of-sight\footnote{Some features are visible in our Gaia16apd spectrum as well, but the features are weaker and the overall S/N is lower, preventing reliable chemical analysis.}. Our medium resolution far-UV spectra (FWHM$\sim120$\,km\,s$^{-1}$) clearly detected and resolved two sets of absorption systems, with one component matching the redshift of the host galaxy NGC\,3191, and the second, $235$\,km\,s$^{-1}$ blue shifted component, consistent with the redshift of the companion galaxy, SDSSJ101857.98+462714.6 (45\,kpc away). We hypothesize that these two galaxies are undergoing tidal interaction, and the two absorption systems originate from halo gas surrounding each galaxy. 

Besides narrow metal absorption lines, our data also detected a Lyman-$\alpha$ absorption from the host galaxy with a total column density of $(6.7\pm1.2)\times10^{19}$\,cm$^{-2}$ (sub-DLA)
This is an order of magnitude smaller than the typical HI column density ($10^{21}$\,cm$^{-1}$) measured in the disks of nearby spiral galaxies by the THINGS survey \citep{Walter2008,Leroy2008}, suggesting that SN2017egm is likely to be on the near side of the galaxy, with a fairly low extinction ($\rm E(B-V)$\,=\,0.007) for the supernova.

Although medium spectral resolution prevents us from accurate measurements of metal abundances of interstellar and circumstellar gas, we are able to set the {\color{black} probable lower limits} to metallicities of the two velocity components to $>1~Z_\odot$. 
This is consistent with nebular gas-phase metallicities measured from optical spectra at the SN location (IFU data) and the companion galaxy (SDSS) {\color{black} \citep{Nicholl2017,Bose2017,Chen2017b,Izzo2017}}. 
Combining the low HI column density result with the high metallicity measurement, we can rule out the scenario where SN2017egm is located in a dwarf companion galaxy. We conclude that although it is rare, a SLSN-I event can explode in a metal rich environment.

We acknowledge many useful discussions with Matt Nicholl at Harvard and Janet Chen at Max-Planck-Institut f\"ur Extraterrestrische Physik, in Germany on the measurements of gas-phase metallicities in NGC3191 and the companion galaxy. 
We thank Nadia Blagorodnova for providing example python emcee scripts for our spectral fitting. We also thank Nick Scoville at Caltech for discussions on the HI column density distribution of nearby galaxies.

{\it Facilities:} \facility{HST}.

{\it Software:} Python, Astropy


\begin{thebibliography}{}
\providecommand\natexlab[1]{#1}

\bibitem[{{Alexandroff} {et~al.}(2015){Alexandroff}, {Heckman}, {Borthakur},
  {Overzier}, \& {Leitherer}}]{Alexandroff2015}
{Alexandroff}, R.~M., {Heckman}, T.~M., {Borthakur}, S., {Overzier}, R., \&
  {Leitherer}, C. 2015,
  \href{http://dx.doi.org/10.1088/0004-637X/810/2/104}{{\apj},
  810, 104}

\bibitem[{{Asplund} {et~al.}(2009){Asplund}, {Grevesse}, {Sauval}, \&
  {Scott}}]{Asplund2009}
{Asplund}, M., {Grevesse}, N., {Sauval}, A.~J., \& {Scott}, P. 2009,
  \href{http://dx.doi.org/10.1146/annurev.astro.46.060407.145222}{{\araa},
  47, 481}

\bibitem[{{Bose} {et~al.}(2017){Bose}, {Dong}, {Pastorello}, {Filippenko},
  {Kochanek}, {Mauerhan}, {Romero-Canizales}, {Brink}, {Chen}, {Prieto},
  {Post}, {Ashall}, {Grupe}, {Tomasella}, {Benetti}, {Shappee}, {Stanek},
  {Cai}, {Falco}, {Lundqvist}, {Mattila}, {Mutel}, {Ochner}, {Pooley},
  {Stritzinger}, {Villanueva}, {Zheng}, {Beswick}, {Brown}, {Cappellaro},
  {Davis}, {de Jaeger}, {Elias-Rosa}, {Gall}, {Gaudi}, {Herczeg}, {Hestenes},
  {Holoien}, {Hosseinzadeh}, {Hsiao}, {Hu}, {Jaejin}, {Jeffers}, {Koff},
  {Kumar}, {Kurtenkov}, {Lau}, {Prentice}, {Rudy}, {Shahbandeh}, {Somero},
  {Stassun}, {Thompson}, {Valenti}, {Woo}, \& {Yunus}}]{Bose2017}
{Bose}, S., {Dong}, S., {Pastorello}, A., {et~al.} 2017, {ArXiv
  e-prints}, \href{http://arxiv.org/abs/1708.00864}{{\sffamily arXiv:1708.00864
  [astro-ph.HE]}}

\bibitem[{{Bufano} {et~al.}(2009){Bufano}, {Immler}, {Turatto}, {Landsman},
  {Brown}, {Benetti}, {Cappellaro}, {Holland}, {Mazzali}, {Milne}, {Panagia},
  {Pian}, {Roming}, {Zampieri}, {Breeveld}, \& {Gehrels}}]{Bufano2009}
{Bufano}, F., {Immler}, S., {Turatto}, M., {et~al.} 2009,
  \href{http://dx.doi.org/10.1088/0004-637X/700/2/1456}{{\apj},
  700, 1456}

\bibitem[{{Chen} {et~al.}(2017{\natexlab{a}}){Chen}, {Smartt}, {Yates},
  {Nicholl}, {Kr{\"u}hler}, {Schady}, {Dennefeld}, \& {Inserra}}]{Chen2017}
{Chen}, T.-W., {Smartt}, S.~J., {Yates}, R.~M., {et~al.} 2017{\natexlab{a}},
  \href{http://dx.doi.org/10.1093/mnras/stx1428}{{\mnras}, 470,
  3566}

\bibitem[{{Chen} {et~al.}(2017{\natexlab{b}}){Chen}, {Schady}, {Xiao},
  {Eldridge}, {Schweyer}, {Lee}, {Yu}, {Smartt}, \& {Inserra}}]{Chen2017b}
{Chen}, T.-W., {Schady}, P., {Xiao}, L., {et~al.} 2017{\natexlab{b}},
  {ArXiv e-prints},
  \href{http://arxiv.org/abs/1708.04618}{{\sffamily arXiv:1708.04618}}

\bibitem[Chomiuk et al.(2011)]{Chomiuk2011} Chomiuk, L., Chornock, R., Soderberg, A.~M., et al.\ 2011, \apj, 743, 114 


\bibitem[{{De Cia} {et~al.}(2016){De Cia}, {Ledoux}, {Mattsson}, {Petitjean},
  {Srianand}, {Gavignaud}, \& {Jenkins}}]{deCia2016}
{De Cia}, A., {Ledoux}, C., {Mattsson}, L., {et~al.} 2016,
  \href{http://dx.doi.org/10.1051/0004-6361/201527895}{{\aap},
  596, A97}

\bibitem[{{De Cia} {et~al.}(2017){De Cia}, {Ledoux}, {Petitjean}, \&
  {Savaglio}}]{deCia2017}
{De Cia}, A., {Ledoux}, C., {Petitjean}, P., \& {Savaglio}, S. 2017,
  {ArXiv e-prints},
  \href{http://arxiv.org/abs/1709.06578}{{\sffamily arXiv:1709.06578}}
\bibitem[De Cia et al.(2017b)]{DeCia2017b} De Cia, A., Gal-Yam, A., Rubin, A., et al.\ 2017, arXiv:1708.01623 


\bibitem[{{Delgado} {et~al.}(2017){Delgado}, {Harrison}, {Hodgkin}, {Leeuwen},
  {Rixon}, \& {Yoldas}}]{Delgado2017}
{Delgado}, A., {Harrison}, D., {Hodgkin}, S., {et~al.} 2017,
  {Transient Name Server Discovery Report}, 591

\bibitem[{{Dessart} {et~al.}(2012){Dessart}, {Hillier}, {Waldman}, {Livne}, \&
  {Blondin}}]{Dessart2012}
{Dessart}, L., {Hillier}, D.~J., {Waldman}, R., {Livne}, E., \& {Blondin}, S.
  2012,
  \href{http://dx.doi.org/10.1111/j.1745-3933.2012.01329.x}{{\mnras},
  426, L76}

\bibitem[{{Dong} {et~al.}(2017){Dong}, {Bose}, {Chen}, {Benetti}, {Pastorello},
  {Cai}, {Prieto}, {Brown}, {Ochner}, {Ashall}, {Stritzinger}, {Lundqvist},
  {Mattila}, {Elias-Rosa}, {Post}, {}, {Stanek}, {Koff}, {NUTS Team}, \&
  {ASAS-SN Collaboration}}]{Dong2017}
{Dong}, S., {Bose}, S., {Chen}, P., {et~al.} 2017, {The
  Astronomer's Telegram}, 1049

\bibitem[Foreman-Mackey et al.(2013)]{Foreman-Mackey2012} Foreman-Mackey, D., Hogg, D.~W., Lang, D., \& Goodman, J.\ 2013, \pasp, 125, 306 

\bibitem[{{Gaia Collaboration} {et~al.}(2016){Gaia Collaboration}, {Prusti},
  {de Bruijne}, {Brown}, {Vallenari}, {Babusiaux}, {Bailer-Jones}, {Bastian},
  {Biermann}, {Evans}, \& et~al.}]{Gaia2016}
{Gaia Collaboration}, {Prusti}, T., {de Bruijne}, J.~H.~J., {et~al.} 2016,
  \href{http://dx.doi.org/10.1051/0004-6361/201629272}{{\aap},
  595, A1}

\bibitem[{{Grimes} {et~al.}(2009){Grimes}, {Heckman}, {Aloisi}, {Calzetti},
  {Leitherer}, {Martin}, {Meurer}, {Sembach}, \& {Strickland}}]{Grimes2009}
{Grimes}, J.~P., {Heckman}, T., {Aloisi}, A., {et~al.} 2009,
  \href{http://dx.doi.org/10.1088/0067-0049/181/1/272}{{\apjs},
  181, 272}

\bibitem[{{Guillochon} {et~al.}(2017){Guillochon}, {Parrent}, {Kelley}, \&
  {Margutti}}]{Guillochon2017}
{Guillochon}, J., {Parrent}, J., {Kelley}, L.~Z., \& {Margutti}, R. 2017,
  \href{http://dx.doi.org/10.3847/1538-4357/835/1/64}{{\apj}, 835,
  64}

\bibitem[{{G{\"u}ver} \& {{\"O}zel}(2009)}]{Guever2009}
{G{\"u}ver}, T., \& {{\"O}zel}, F. 2009,
  \href{http://dx.doi.org/10.1111/j.1365-2966.2009.15598.x}{{\mnras},
  400, 2050}

\bibitem[{{Heckman} {et~al.}(2015){Heckman}, {Alexandroff}, {Borthakur},
  {Overzier}, \& {Leitherer}}]{Heckman2015}
{Heckman}, T.~M., {Alexandroff}, R.~M., {Borthakur}, S., {Overzier}, R., \&
  {Leitherer}, C. 2015,
  \href{http://dx.doi.org/10.1088/0004-637X/809/2/147}{{\apj},
  809, 147}

\bibitem[{{Howell} {et~al.}(2013){Howell}, {Kasen}, {Lidman}, {Sullivan},
  {Conley}, {Astier}, {Balland}, {Carlberg}, {Fouchez}, {Guy}, {Hardin},
  {Pain}, {Palanque-Delabrouille}, {Perrett}, {Pritchet}, {Regnault}, {Rich},
  \& {Ruhlmann-Kleider}}]{Howell2013}
{Howell}, D.~A., {Kasen}, D., {Lidman}, C., {et~al.} 2013,
  \href{http://dx.doi.org/10.1088/0004-637X/779/2/98}{{\apj}, 779,
  98}

\bibitem[{{Izzo} {et~al.}(2017){Izzo}, {Th{\"o}ne}, {Garc{\'{\i}}a-Benito}, {de
  Ugarte Postigo}, {Cano}, {Kann}, {Bensch}, {Galad{\'{\i}}-Enr{\'{\i}}quez},
  {Hedrosa}, \& {Della Valle}}]{Izzo2017}
{Izzo}, L., {Th{\"o}ne}, C.~C., {Garc{\'{\i}}a-Benito}, R., {et~al.} 2017,
  {ArXiv e-prints},
  \href{http://arxiv.org/abs/1708.03856}{{\sffamily arXiv:1708.03856
  [astro-ph.HE]}}

\bibitem[{{Leloudas} {et~al.}(2015){Leloudas}, {Schulze}, {Kr{\"u}hler},
  {Gorosabel}, {Christensen}, {Mehner}, {de Ugarte Postigo}, {Amor{\'{\i}}n},
  {Th{\"o}ne}, {Anderson}, {Bauer}, {Gallazzi}, {He{\l}miniak}, {Hjorth},
  {Ibar}, {Malesani}, {Morell}, {Vinko}, \& {Wheeler}}]{Leloudas2015}
{Leloudas}, G., {Schulze}, S., {Kr{\"u}hler}, T., {et~al.} 2015,
  \href{http://dx.doi.org/10.1093/mnras/stv320}{{\mnras}, 449,
  917}

\bibitem[{{Leroy} {et~al.}(2008){Leroy}, {Walter}, {Brinks}, {Bigiel}, {de
  Blok}, {Madore}, \& {Thornley}}]{Leroy2008}
{Leroy}, A.~K., {Walter}, F., {Brinks}, E., {et~al.} 2008,
  \href{http://dx.doi.org/10.1088/0004-6256/136/6/2782}{{\aj},
  136, 2782}

\bibitem[{{Lunnan} {et~al.}(2014){Lunnan}, {Chornock}, {Berger}, {Laskar},
  {Fong}, {Rest}, {Sanders}, {Challis}, {Drout}, {Foley}, {Huber}, {Kirshner},
  {Leibler}, {Marion}, {McCrum}, {Milisavljevic}, {Narayan}, {Scolnic},
  {Smartt}, {Smith}, {Soderberg}, {Tonry}, {Burgett}, {Chambers}, {Flewelling},
  {Hodapp}, {Kaiser}, {Magnier}, {Price}, \& {Wainscoat}}]{Lunnan2014}
{Lunnan}, R., {Chornock}, R., {Berger}, E., {et~al.} 2014,
  \href{http://dx.doi.org/10.1088/0004-637X/787/2/138}{{\apj},
  787, 138}

\bibitem[{{Lunnan} {et~al.}(2017){Lunnan}, {Chornock}, {Berger}, {Jones},
  {Rest}, {Czekala}, {Dittmann}, {Drout}, {Foley}, {Fong}, {Kirshner},
  {Laskar}, {Leibler}, {Margutti}, {Milisavljevic}, {Narayan}, {Pan}, {Riess},
  {Roth}, {Sanders}, {Scolnic}, {Smartt}, {Smith}, {Chambers}, {Draper},
  {Flewelling}, {Huber}, {Kaiser}, {Kudritzki}, {Magnier}, {Metcalfe},
  {Wainscoat}, {Waters}, \& {Willman}}]{Lunnan2017}
---. 2017, {ArXiv e-prints},
  \href{http://arxiv.org/abs/1708.01619}{{\sffamily arXiv:1708.01619
  [astro-ph.HE]}}
\bibitem[Marasco \& Fraternali(2011)]{Marasco2011} Marasco, A., \& Fraternali, F.\ 2011, \aap, 525, A134 


\bibitem[{{Mazzali} {et~al.}(2016){Mazzali}, {Sullivan}, {Pian}, {Greiner}, \&
  {Kann}}]{Mazzali2016}
{Mazzali}, P.~A., {Sullivan}, M., {Pian}, E., {Greiner}, J., \& {Kann}, D.~A.
  2016, \href{http://dx.doi.org/10.1093/mnras/stw512}{{\mnras},
  458, 3455}

\bibitem[Nakar \& Sari(2010)]{Nakar2010} Nakar, E., \& Sari, R.\ 2010, \apj, 725, 904 


\bibitem[Nicholl et al.(2013)]{Nicholl2013} Nicholl, M., Smartt, S.~J., Jerkstrand, A., et al.\ 2013, \nat, 502, 346 

\bibitem[Nicholl et al.(2015)]{Nicholl2015} Nicholl, M., Smartt, S.~J., Jerkstrand, A., et al.\ 2015, \mnras, 452, 3869 


\bibitem[{{Nicholl} {et~al.}(2017{\natexlab{a}}){Nicholl}, {Berger},
  {Margutti}, {Blanchard}, {Guillochon}, {Leja}, \& {Chornock}}]{Nicholl2017}
{Nicholl}, M., {Berger}, E., {Margutti}, R., {et~al.} 2017{\natexlab{a}},
  {ArXiv e-prints},
  \href{http://arxiv.org/abs/1706.08517}{{\sffamily arXiv:1706.08517
  [astro-ph.HE]}}

\bibitem[{{Nicholl} {et~al.}(2017{\natexlab{c}}){Nicholl}, {Guillochon}, \&
  {Berger}}]{Nicholl2017b}
{Nicholl}, M., {Guillochon}, J., \& {Berger}, E. 2017{\natexlab{c}},
  {ArXiv e-prints},
  \href{http://arxiv.org/abs/1706.00825}{{\sffamily arXiv:1706.00825
  [astro-ph.HE]}}

\bibitem[{{Nicholl} {et~al.}(2017{\natexlab{b}}){Nicholl}, {Berger},
  {Margutti}, {Blanchard}, {Milisavljevic}, {Challis}, {Metzger}, \&
  {Chornock}}]{Nicholl2017c}
---. 2017{\natexlab{b}},
  \href{http://dx.doi.org/10.3847/2041-8213/aa56c5}{{\apjl}, 835,
  L8}

\bibitem[{{Pauldrach} {et~al.}(1996){Pauldrach}, {Duschinger}, {Mazzali},
  {Puls}, {Lennon}, \& {Miller}}]{Pauldrach1996}
{Pauldrach}, A.~W.~A., {Duschinger}, M., {Mazzali}, P.~A., {et~al.} 1996,
  {\aap}, 312, 525

\bibitem[{{Perley} {et~al.}(2016){Perley}, {Quimby}, {Yan}, {Vreeswijk}, {De
  Cia}, {Lunnan}, {Gal-Yam}, {Yaron}, {Filippenko}, {Graham}, {Laher}, \&
  {Nugent}}]{Perley2016}
{Perley}, D.~A., {Quimby}, R.~M., {Yan}, L., {et~al.} 2016,
  \href{http://dx.doi.org/10.3847/0004-637X/830/1/13}{{\apj}, 830,
  13}

\bibitem[{{P{\'e}roux} {et~al.}(2003){P{\'e}roux}, {Dessauges-Zavadsky},
  {D'Odorico}, {Kim}, \& {McMahon}}]{Peroux2003}
{P{\'e}roux}, C., {Dessauges-Zavadsky}, M., {D'Odorico}, S., {Kim}, T.-S., \&
  {McMahon}, R.~G. 2003,
  \href{http://dx.doi.org/10.1046/j.1365-8711.2003.06952.x}{{\mnras},
  345, 480}

\bibitem[{{Pettini} \& {Pagel}(2004)}]{Pettini2004}
{Pettini}, M., \& {Pagel}, B.~E.~J. 2004,
  \href{http://dx.doi.org/10.1111/j.1365-2966.2004.07591.x}{{\mnras},
  348, L59}

\bibitem[{{Quimby} {et~al.}(2011){Quimby}, {Kulkarni}, {Kasliwal}, {Gal-Yam},
  {Arcavi}, {Sullivan}, {Nugent}, {Thomas}, {Howell}, {Nakar}, {Bildsten},
  {Theissen}, {Law}, {Dekany}, {Rahmer}, {Hale}, {Smith}, {Ofek}, {Zolkower},
  {Velur}, {Walters}, {Henning}, {Bui}, {McKenna}, {Poznanski}, {Cenko}, \&
  {Levitan}}]{Quimby2011}
{Quimby}, R.~M., {Kulkarni}, S.~R., {Kasliwal}, M.~M., {et~al.} 2011,
  \href{http://dx.doi.org/10.1038/nature10095}{{\nat}, 474, 487}

\bibitem[{{Schulze} {et~al.}(2016){Schulze}, {Kr{\"u}hler}, {Leloudas},
  {Gorosabel}, {Mehner}, {Buchner}, {Kim}, {Ibar}, {Amor{\'{\i}}n},
  {Herrero-Illana}, {Anderson}, {Bauer}, {Christensen}, {de Pasquale}, {de
  Ugarte Postigo}, {Gallazzi}, {Hjorth}, {Morrell}, {Malesani}, {Sparre},
  {Stalder}, {Stark}, {Th{\"o}ne}, \& {Wheeler}}]{Schulze2016}
{Schulze}, S., {Kr{\"u}hler}, T., {Leloudas}, G., {et~al.} 2016,
  {ArXiv e-prints},
  \href{http://arxiv.org/abs/1612.05978}{{\sffamily arXiv:1612.05978}}

\bibitem[{{Thomas} {et~al.}(2011){Thomas}, {Nugent}, \& {Meza}}]{Thomas2011}
{Thomas}, R.~C., {Nugent}, P.~E., \& {Meza}, J.~C. 2011,
  \href{http://dx.doi.org/10.1086/658673}{{\pasp}, 123, 237}

\bibitem[{{Vreeswijk} {et~al.}(2014){Vreeswijk}, {Savaglio}, {Gal-Yam}, {De
  Cia}, {Quimby}, {Sullivan}, {Cenko}, {Perley}, {Filippenko}, {Clubb},
  {Taddia}, {Sollerman}, {Leloudas}, {Arcavi}, {Rubin}, {Kasliwal}, {Cao},
  {Yaron}, {Tal}, {Ofek}, {Capone}, {Kutyrev}, {Toy}, {Nugent}, {Laher},
  {Surace}, \& {Kulkarni}}]{Vreeswijk2014}
{Vreeswijk}, P.~M., {Savaglio}, S., {Gal-Yam}, A., {et~al.} 2014,
  \href{http://dx.doi.org/10.1088/0004-637X/797/1/24}{{\apj}, 797,
  24}

\bibitem[Vreeswijk et al.(2017)]{Vreeswijk2017} Vreeswijk, P.~M., Leloudas, G., Gal-Yam, A., et al.\ 2017, \apj, 835, 58 

\bibitem[{{Walter} {et~al.}(2008){Walter}, {Brinks}, {de Blok}, {Bigiel},
  {Kennicutt}, {Thornley}, \& {Leroy}}]{Walter2008}
{Walter}, F., {Brinks}, E., {de Blok}, W.~J.~G., {et~al.} 2008,
  \href{http://dx.doi.org/10.1088/0004-6256/136/6/2563}{{\aj},
  136, 2563}

\bibitem[{{Werk} {et~al.}(2014){Werk}, {Prochaska}, {Tumlinson}, {Peeples},
  {Tripp}, {Fox}, {Lehner}, {Thom}, {O'Meara}, {Ford}, {Bordoloi}, {Katz},
  {Tejos}, {Oppenheimer}, {Dav{\'e}}, \& {Weinberg}}]{Werk2014}
{Werk}, J.~K., {Prochaska}, J.~X., {Tumlinson}, J., {et~al.} 2014,
  \href{http://dx.doi.org/10.1088/0004-637X/792/1/8}{{\apj}, 792,
  8}

\bibitem[{{Yan} {et~al.}(2017){Yan}, {Quimby}, {Gal-Yam}, {Brown},
  {Blagorodnova}, {Ofek}, {Lunnan}, {Cooke}, {Cenko}, {Jencson}, \&
  {Kasliwal}}]{Yan2017}
{Yan}, L., {Quimby}, R., {Gal-Yam}, A., {et~al.} 2017,
  \href{http://dx.doi.org/10.3847/1538-4357/aa6b02}{{\apj}, 840,
  57}

\end{thebibliography}

\begin{deluxetable*}{cccccccc}
\tablecolumns{8}
\tablewidth{0pc}
\tabletypesize{\scriptsize}
\tablecaption{\label{obslog} {\it HST/UV} Spectroscopy Observation Log }
\tablehead{\colhead{Obs.UT} & \colhead{Name} & \colhead{Exp.Time} & \colhead{Instrument} & Grating & \colhead{$\Delta \lambda$}  & \colhead{Spec Resolution} & \colhead{Obs.setup} \\
                                               &                                 &                                    &               & \colhead{\AA}   &   &  \\
 }
 \startdata
2017-06-23 & SN2017egm & 7488.9  & COS/FUV$^a$   & G140L & 1118 - 2251  & 1500-4000 & Seg-A/Time-TAG \\
2017-06-23 & SN2017egm & 1681.2  & STIS/NUV      & G230L & 1570 - 3180  & 500-1010  & NUV-MAMA \\ 
2017-06-23 & SN2017egm & 300.0   & STIS/NUV      & G430L & 2900 - 5700  & 500-1010  & NUV-MAMA \\
2012-06-02 & Gaia16apd & 4889    & COS/FUV       & G140L & 1118 - 2251  & 1500-4000 &Seg-A/Time-TAG \\
2016-06-02 & Gaia16apd & 2327    & STIS/NUV      & G230L & 1570 - 3180  & 500-1010  & NUV-MAMA \\
2012-05-26 & PTF12dam  & 2448    & WFC3/UVIS     & G280  & 1840 - 4500  & 70 @3000\AA & UVIS \\

 \enddata
 \tablenotetext{a}{COS/FUV data was taken using only Segment A.}
 \end{deluxetable*}

\begin{deluxetable}{cccccc}
\tablecolumns{6}
\tablewidth{0pc}
\tabletypesize{\scriptsize}
\tablecaption{\label{tab:date} Phase information for the four SLSNe-I}
\tablehead{\colhead{Name} & \colhead{Redshift} & \colhead{Spec.Date} & \colhead{Peak Date} & \colhead{Exp. Date}  & \colhead{$\Delta t_{exp}$} \\
                        &                    &          Day        &            Day   &    Day                                 &    Day           \\
}
\startdata
SN2017egm & 0.03   & 57927 & 57924 & 57890    &  35.9   \\
Gaia16apd & 0.1018 & 57541 & 57541 & 57505    &  32.7   \\
PTF12dam  & 0.1078 & 56073 & 56096.7 & 56020.9 & 47   \\
iPTF13ajg  & 0.7403 & 56399 & 56405 & 56350    &  28.2 \\
\enddata
\end{deluxetable}

\begin{deluxetable}{ccccc}
\tablecolumns{5}
\tablewidth{0pc}
\tabletypesize{\scriptsize}
\tablecaption{\label{mcmcoutput} Results from the MCMC simulation - variable $\lambda_0$}
\tablehead{\colhead{Obj.} & \colhead{T} & \colhead{$\beta$} & \colhead{$\lambda_0$} & \colhead{$\log_{10}(A)$} \\
                                          &  K                &                              & $\AA$                            &                                      \\
}
\startdata
SN2017egm  &  $17095^{+462}_{-193}$  & $1.95^{+0.10}_{-0.09}$ & $2817.5^{+78}_{-10}$ & $8.79^{+0.02}_{-0.01}$ \\
Gaia16apd  & $17403^{+3131}_{-647}$   & $0.51^{+0.4}_{-0.16}$ & $3133^{+453}_{-549}$  & $8.01^{+0.06}_{-0.04}$ \\
iPTF13ajg  & $15162^{+254}_{-182}$    & $2.86^{+0.18}_{-0.09}$ & $2820^{+13}_{-10}$  & $6.18^{+0.02}_{-0.02}$ \\
PTF12dam   & $14681^{+330}_{-141}$    & $1.78^{+0.08}_{-0.06}$ & $2819^{+51}_{-17}$  & $10.98^{+0.26}_{-0.01}$ \\

\enddata
\end{deluxetable}

\begin{deluxetable}{cccc}
\tablecolumns{4}
\tablewidth{0pc}
\tabletypesize{\scriptsize}
\tablecaption{\label{mcmc2} Results from the MCMC simulation - fixed $\lambda_0$}
\tablehead{\colhead{Obj.} & \colhead{T} & \colhead{$\beta$} & \colhead{$\log_10(A)$} \\
                                          &  K                &                              &                                      \\
}
\startdata
SN2017egm  &  $17094^{+587}_{-285}$  & $1.96^{+0.15}_{-0.14}$ & $8.80^{+0.03}_{-0.02}$ \\
Gaia16apd  & $16703^{+480}_{-398}$   & $0.80^{+0.15}_{-0.16}$ & $8.04^{+0.05}_{-0.02}$ \\
iPTF13ajg  & $15008^{+200}_{-75}$   & $2.94^{+0.09}_{-0.04}$ &  $6.19^{+0.01}_{-0.01}$ \\
PTF12dam   & $14681^{+88}_{-76}$    & $1.78^{+0.03}_{-0.03}$ &  $10.98^{+0.03}_{-0.01}$ \\

\enddata
\end{deluxetable}

\begin{deluxetable}{lccr}
\tablecolumns{4}
\tablewidth{0pc}
\tabletypesize{\scriptsize}
\tablecaption{\label{tab:line} The Measured Column Densities and Abundances}
\tablehead{ \colhead{  } & \colhead{Component 1} & \colhead{Component 2} & \colhead{Total}  \\ 
\\
                                       &   at z=0.0307                 &  at z=0.0299                   &                  \\  }
\startdata
N(SII)$^a$  & $(8.10\pm1.90)$e+14   & $(1.20\pm0.18)$e+15       & $(2.01\pm0.26)$e+15   \\
N(FeII)     & $(2.13\pm0.21)$e+14 & $(4.80\pm1.60)$e+13        & $(2.61\pm0.26)$e+14  \\
N(AlII)     & $(6.31\pm0.65)$e+12 & $(3.64\pm0.56)$e+12      & $(9.95\pm0.86)$e+12  \\ 
N(HI)       & $(3.00\pm0.80)$e+19   & $(3.70\pm0.90)$e+19        & $(6.70\pm1.20)$e+19    \\
\hline \\
$\rm [S/H]_{obs}^b$ & $+0.30\pm 0.16$ & $+0.38\pm 0.12$ & $+0.34\pm 0.10 $  \\
$\rm [Fe/H]_{obs}$  & $-0.62\pm 0.12$ & $-1.36\pm 0.18$ & $-0.88\pm 0.09 $  \\
$\rm [Al/H]_{obs}$  & $-1.12\pm 0.12$ & $-1.45\pm 0.13$ & $-1.27\pm 0.09 $  \\
$\rm [S/Fe]_{obs}$  & $+0.92\pm 0.11$ & $+1.74\pm 0.16$ & $+1.23\pm 0.13 $  \\
$\rm [S/Al]_{obs}$  & $+1.42\pm 0.11$ & $+1.83\pm 0.09$ & $+1.62\pm 0.10 $  \\
\hline \\
$\rm [S/H]_{DC}^c$ & $+0.59\pm 0.16$ & $+0.98\pm 0.12$ & $+0.74\pm 0.10 $ \\
$\rm [Fe/H]_{DC}$  & $+0.48\pm 0.12$ & $+1.10\pm 0.18$ & $+0.74\pm 0.09 $ \\
$\rm [Al/H]_{DC}$  & $-0.01\pm0.12$  & $+1.01\pm 0.13$ & $+0.34\pm 0.09 $ \\
\hline \\
$\rm [S/H]_{HC}^c$& $+0.06\pm 0.16$  & $+0.14\pm 0.12$ & $+0.11\pm 0.10 $ \\
$\rm [Fe/H]_{HC}$ & $-0.86\pm 0.12$  & $-1.60\pm 0.18$ & $-1.12\pm 0.09 $ \\
$\rm [Al/H]_{HC}$ & $-1.35\pm 0.12$  & $-1.68\pm 0.13$ & $-1.50\pm 0.09 $ \\
\hline \\
$\rm [S/H]_{corr}^c$ &+$0.35\pm0.16$  & $+0.75\pm0.12$ &+$0.51\pm0.10$ \\
$\rm [Fe/H]_{corr}$  &+$0.25\pm0.12$  & $+0.87\pm0.18$ &+$0.51\pm0.09$ \\
$\rm [Al/H]_{corr}$  &$-0.25\pm0.12$  & $+0.78\pm0.13$ &+$0.11\pm0.09$ \\

\enddata
\tablenotetext{a}{Column density N(X) for ion X is in units of cm$^{-2}$.}
\tablenotetext{b}{Abundance [X/H] is in logrithmic and relative to solar abundance for ion X. It is defined as 12 + $\log_{10}{(\rm N(X)/N(HI)) - [X/H]_\odot}$. Here [S/H]$_\odot = 7.135$, [Fe/H]$_\odot = 7.475$ and [Al/H]$_\odot = 6.44$ \citep{Asplund2009}.}
\tablenotetext{c}{[X/H]$_{\rm obs}$ is the value computed from the observed column densities without any correction. [X/H]$_{\rm DC}$ is for dust depletion corrected value,  [X/H]$_{\rm HC}$ stands for the values included H ionization correction, and [X/H]$_{\rm corr}$ is the abundance including both dust and H ionization correction.}
\end{deluxetable}

\end{document}